\documentclass[11pt]{amsart}

\usepackage{graphicx, color, appendix, url,tikz,hyperref} % Required for inserting images
\usetikzlibrary{positioning}

\usepackage{amsfonts}
\usepackage{amsmath}
\usepackage{amsthm}
\usepackage{amssymb}

\usepackage{tabularx}

\usepackage[left=1in,right=1in,top=1in,bottom=1in]{geometry}

\newtheorem{theorem}{Theorem}[section]

\newtheorem*{theorem*}{Theorem}

\theoremstyle{definition}

\newtheorem{example}[theorem]{Example}

\theoremstyle{remark}

\newtheorem{question}[theorem]{Question}

\begin{document}

%\begin{frontmatter}

\title{Structural identifiability of compartmental models: \\ Recent progress and future directions}

\author{Nicolette Meshkat}
\author{Anne Shiu}
%\author[1]{Nicolette Meshkat}
%\author[2]{Anne Shiu} %\corref{cor1}%

%\affiliation[1]{organization={Santa Clara University}}
%\affiliation[2]{organization={Texas A\&M University}}

\date{\today}

\begin{abstract} 
We summarize recent progress on the theory and applications of structural identifiability of compartmental models.   
On the applications side, we review identifiability analyses undertaken recently for models arising in epidemiology, oncology, and other areas; and we summarize common approaches for handling models that are unidentifiable. 
We also highlight recent theoretical and algorithmic results on how to reparametrize unidentifiable models and, in the context of linear compartmental models, how to predict identifiability properties directly from the model structure. 
Finally, we highlight future research directions.

%\vspace{.1in}
%\noindent
% {\bf Keywords:} Linear compartmental model, structural identifiability, 
\end{abstract}

%\begin{keyword}
%    Compartmental model \sep structural identifiability \sep reparametrization
% {\color{red}[provide 1 to 7 keywords for indexing purposes]}
%\end{keyword}

%\end{frontmatter}
\maketitle

\section{Introduction}
Parameter estimation is a key step in calibrating a mathematical model, but before doing so we must know whether the model structure enables the parameters to be recovered from experimental data.  Accordingly, we say that a model is (generically) locally {\em structurally identifiable} -- or {\em identifiable}, for short -- if (generic) values of the parameters can be recovered, up to a finite set, from 
noiseless experimental data.  Structural identifiability is necessary for {\em practical identifiability}, which allows for experimental data with noise. %~\cite{Tuncer-Le}. 

Recent years have seen renewed interest in the problem of structural identifiability in the context of compartmental models.  This interest comes from both the theory side as well as from applications, many of which come from systems biology.  This article aims to provide an overview of recent results, and additionally highlight open research questions. % inspired by recent progress. 

We focus on three main topics. 
The first topic is in the direction of applications (Section~\ref{sec:applications}).  We review recent identifiability analyses of models arising in epidemiology, cancer modeling, and other areas; in particular, we summarize common approaches used to handle models that are unidentifiable.
%(in addition to the reparametrization approach mentioned above).  
%
Our second topic continues with the theme of dealing with unidentifiable models, this time from the theoretical and algorithmic perspective (Section~\ref{sec:reparam}).  Specifically, we review recent approaches to reparametrizing models so that the resulting models are identifiable. 
Our third topic is on the identifiability of {\em linear compartmental models} -- in which the ODEs are linear and the model is summarized by a directed graph.
Even in this setting there is no complete answer to the question of which models are identifiable (to say nothing about the identifiability of individual parameters).  Nevertheless, significant partial answers have been given in recent years, which we highlight in Section~\ref{sec:iden-model}.

\subsection{Other recent advances}
Progress on the structural identifiability problem
has been driven in part by improved methods -- algorithms and software -- for assessing identifiability. 
%in theory, software, and applications.
Here we briefly mention a few of these advances.
%, but in lesser detail than the topics mentioned above which are the focus of this review.
New algorithms for testing global identifiability 
%and 
have come from advances in theory~\cite{OPT,HOPY, multi-experiment,joubert, borgqvist,massonis2020finding}, leading to new software \cite{julia-software, sian, ilmer2021web, stigter2021computing}. 
An analysis of the available software was conducted recently
%A recent benchmarking comparison analysis of the available software can be found in
\cite{rey2023benchmarking}, and software tutorials can be found \cite{tutorial-identifiability-julia, chowell2023structural, beginner-guide}. 
Various methods have also been extended
to new areas, such as PDEs \cite{browningPDE, pde-byrne, liu, salmaniw2024structural}, stochastic models~\cite{browning, janzen2019three}, and machine learning \cite{norden2025importance}.  
Finally, the availability of new user-friendly software has enabled the construction of new databases of linear compartmental models and their identifiability properties \cite{gogishvili-database} (see also~\cite[Appendix~B]{cycle-cat}). 

\section{Background} \label{sec:background}
This section recalls the concepts of identifiability and compartmental models.

\subsection{Structural identifiability} \label{sec:iden-defn}
We consider a common experimental setup that consists of a model given by a system of ordinary differential equations (ODEs) as follows:
    \begin{align}
        \label{eq:ODE}
        \dot x ~&=~ f(x(t), \alpha , u(t)) \\
        \notag
        y(t) ~&=~ g(x(t),\alpha ))~, 
    \end{align}
where $x(t) \in \mathbb{R}^N$ denotes the vector of state variables, $\alpha \in \mathbb{R}^n$ is a vector of unknown parameters, $u(t)$ is an input vector (which can be controlled by the experimenter), and $y(t)$ is the output vector.  The model~\eqref{eq:ODE} is (structurally) {\em generically globally identifiable} (respectively, {\em generically locally identifiable}) if from a generic
choice of the input vector $u(t)$, parameter vector $\alpha$, and initial conditions $x(0)$, the parameter vector $\alpha$ can be recovered (respectively, recovered up to a finite set) from perfect measurements of both the input and output vectors.  A model that is not generically locally identifiable is {\em unidentifiable}.  

Similarly, an individual parameter $\alpha_i$ is {\em generically globally identifiable} (respectively, {\em generically locally identifiable} or {\em unidentifiable}) if its value can be recovered (respectively, recovered up to a finite set or can not be recovered even up to a finite set) in the same generic context. We use the acronym {\em SLING} to refer to parameters that are (structurally) locally identifiable but not globally~\cite{sling}.

\begin{example}[Identifiable model]\label{ex:cycle-4-first-instance}
The following model is generically locally identifiable~\cite{cycle-cat}:
\begin{align} \label{eq:cycle}
\dot x_1 ~&=~ a_{14} x_4(t) + (-a_{01} - a_{21}) x_1(t) + u_1(t) \\
    \notag
\dot x_2 ~&=~ a_{21} x_1(t) - a_{32} x_2(t) \\
    \notag
\dot x_3 ~&=~ a_{32} x_2(t) + (-a_{03} - a_{43})x_3(t)  \\
    \notag
\dot x_4 ~&=~ -a_{14} x_4(t) + a_{43} x_3(t) \\
    \notag
y_2(t) ~&=~ x_2(t)~.
\end{align}
Additionally, the parameter $a_{21}$ is generically globally identifiable, while the remaining $5$ parameters are SLING (this is readily computed using software, e.g.~\cite{julia-software}).
\end{example}

\subsection{Compartmental models} \label{sec:compartment}
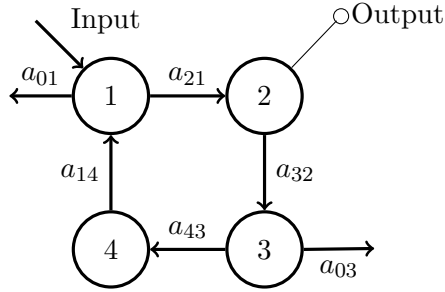
\begin{figure}[ht]
\begin{center}
\begin{tikzpicture}[
roundnode/.style={circle, draw=black, very thick, minimum size=10mm},
arrowbasic/.style={very thick, ->},
]

%Nodes
\node[roundnode](comp1){1};
\node[roundnode](comp2) [right=of comp1] {2};
\node[roundnode](comp3) [below=of comp2] {3};
\node[roundnode](comp4) [left=of comp3] {4};

%Lines

% edges
\draw[arrowbasic] (comp1) -- (comp2) node[pos=.5, above] {\(a_{21}\)};
\draw[arrowbasic] (comp2) -- (comp3) node[pos=.5, right] {\(a_{32}\)};
\draw[arrowbasic] (comp3) -- (comp4) node[pos=.5, above] {\(a_{43}\)};
\draw[arrowbasic] (comp4) -- (comp1) node[pos=.5, left] {\(a_{14}\)};

% input
%\draw[arrowbasic] (0,1.35) -- node[right] {In} (comp1);
\draw[arrowbasic] (-1,1) -- node[above right] {Input} (comp1);

% output
\draw(comp2) -- (3,1);
\draw(3.06,1.06) circle (0.1) node[right] {Output};

% leak
\draw[arrowbasic](comp1) -- node[above] {\(a_{01}\)} (-1.35,0);
%\draw[arrowbasic](comp3) -- node[right] {\(k_{03}\)} (2.05,-3.5);
% LEAK
\draw[arrowbasic] (comp3.east)  -- node[below]{\(a_{03}\)}(3.5, -2.03);
\end{tikzpicture}
\end{center}
    \caption{A linear compartmental model with $In=\{1\}$, $Out=\{2\}$, and $Leak=\{1,3\}$. }
    \label{fig:cycle-4}
\end{figure}

Compartmental models appear frequently in applications, especially in systems biology.  Such models consist of various groups (of individuals, organs, tissue types, etc.) called {\em compartments}, together with the exchanges, interactions, and flows between them.  
An important special case is that of {\em linear compartmental models}, in which the ODE system is linear.  In this setting, the flows between compartments
are represented by a directed graph, with certain nodes representing inputs, outputs, and leaks from the system (we use $In$, $Out$, and $Leak$ to denote these sets).  For instance, the model~\eqref{eq:cycle} in Example~\ref{ex:cycle-4-first-instance} is represented by the directed graph in Figure~\ref{fig:cycle-4}.

% ANOTHER FIGURE
\begin{figure}[ht]
\begin{center}
\begin{tikzpicture}[
roundnode/.style={circle, draw=black, very thick, minimum size=10mm},
arrowbasic/.style={very thick, ->},
]

%Nodes
\node[roundnode](middle){2};
\node[roundnode](leftcomp)  [left=of middle] {1};
\node[roundnode](rightcomp) [below=of middle] {3};

%Lines
\draw[arrowbasic] (leftcomp.east) -- (middle.west) node[pos=.5, above] {\(a_{21}\)};
\draw[arrowbasic] (middle.south east)  -- (rightcomp.north east) node[pos=.5, right] {\(a_{32}\)}; 
\draw[arrowbasic] (rightcomp.north) -- (middle.south) node[pos=.5, left] {\(a_{23}\)};
\draw[arrowbasic] (rightcomp.north west) -- (leftcomp.south east) node[pos=.5, left] {\(a_{13}\)};
 %

%------------------
%======INPUT/OUTPUT/LEAKS 
%------------------
% Input
\draw[arrowbasic] (-2.05,1.35) -- node[right] {Input}(leftcomp.north);
% Output
\draw(-3, 1) circle (0.1) node[left] {Output}; %{$y$};
\draw(-2.94,0.94) -- (leftcomp.north west);
% LEAK
\draw[arrowbasic] (middle.east)  -- node[below]{\(a_{02}\)}(1.5, 0);
% LEAK
\draw[arrowbasic] (rightcomp.east)  -- node[below]{\(a_{03}\)}(1.5, -2.03);
% LEAK
\draw[arrowbasic] (leftcomp.west) -- node[below] {\(a_{01}\)}(-3.5, 0);
\end{tikzpicture}
\end{center}
    \caption{A linear compartmental model in $In=Out=\{1\}$ and $Leak=\{2,3\}$.}
    \label{fig:cycle-3}
\end{figure}
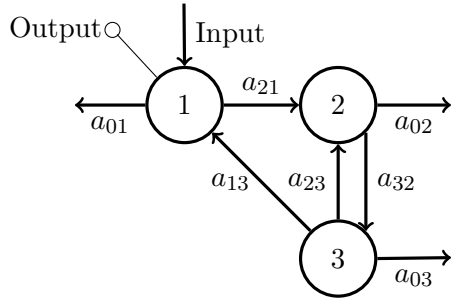

The next example makes use of {\em identifiable functions} of parameters, which  essentially are the functions of the parameters that can be computed from (input-output) data (see e.g.~\cite{all-identifiable-functinos} for a precise definition).

\begin{example}[Unidentifiable model]\label{example:reparam} 
The linear compartmental model shown in Figure~\ref{fig:cycle-3} is a ``single-in-single-out" model~\cite[Example~3]{MOS} given by the following ODEs:
    \begin{align*}
        \dot{x}_1 ~&=~ -a_{01}x_1(t) +a_{13}x_3(t) + u_1(t) \\
        \dot{x}_2 ~&=~ a_{21}x_1(t) -a_{02}x_2(t) + a_{23}x_3(t) \\
        \dot{x}_3 ~&=~ a_{32}x_2(t) -a_{03}x_3(t) \\
        y_1(t) ~&=~ x_1(t)
    \end{align*}
The identifiable functions (which can be computed readily by software, e.g., using~\cite{sian}) are found to be $k_1:=a_{01}$, $k_2:= a_{02}+a_{03}$, $k_3:=a_{13}a_{32}a_{21}$, and $k_4:=a_{02}a_{03}-a_{23}a_{32}$.  We conclude that $a_{01}$ is (globally) identifiable, and it turns out that the remaining $6$ parameters are unidentifiable~\cite{MOS}.  Therefore, this model is structurally unidentifiable.

Next, using the method in \cite{MOS}, we obtain 
an identifiable reparametrization of our model:
\begin{align*}
        \dot{X}_1 ~&=~ X_2 + u_1 \\
        \dot{X}_2 ~&=~ X_3 -k_1 u_1\\
        \dot{X}_3 ~&=~ (-k_1 k_4 + k_3)X_1 - (k_1 k_2 +k_4)X_2 -(k_1 +k_2)X_3 +k_1^2u_1 \\
        y_1 ~&=~ X_1~,
    \end{align*}
where we use the following linear reparametrization:
\begin{align*}
    X_1 ~&=~ x_1 \\
    X_2 ~&=~ -a_{01}x_1 + a_{13}x_3 \\
    X_3 ~&=~ a_{01}^2 x_1 + a_{13}a_{32}x_2 - (a_{01}a_{13}+a_{03}a_{13})x_3~.
\end{align*}
\end{example}

\section{Structural identifiability of (nonlinear) compartmental models in applications}
\label{sec:applications}

%Nonlinear compartmental models are used often in applications; a classic such example is the Kermack and McKendrick SIR Model in epidemiology.  
This section highlights recent applications of compartmental models and their structural identifiability or nonidentifiability.  We begin with applications from epidemiology, which is where compartmental models appear the most frequently (a classic example is the Kermack and McKendrick SIR Model).  
%Compartmental models appear most frequently in epidemiology.  
In many such studies, models are found to be unidentifiable, and subsequently the modelers pinpoint which parameters are unidentifiable and additionally adjust the model in some way so that the updated model is identifiable.  
A summary appears in Table~\ref{tab:applications}, from which we can see that researchers often adjust models by assuming that some initial conditions or some parameter values are known.  Other adjustments include allowing for more observations (that is, adding outputs), and also reparametrizing (e.g., rescaling) or simplifying the model.  We also direct the reader to a recent analysis of 255 variations of Covid transmission models~\cite{massonis2021structural}, which revealed that certain parameters are structurally identifiable in most of the models, but no parameter that is identifiable in all of the models.

\begin{table}[ht]
\caption{{\bf Unidentifiable models in epidemiology.}  This table lists classes of compartmental models analyzed recently and, when relevant, how unidentifiable models were adjusted to become identifiable.
}
\begin{tabularx}{\textwidth}{X X}
    \hline
    Model(s) & Model adjustment(s) \\
    \hline
    \hline
    8 disease models~\cite{chowell2023structural} plus 3 more~\cite{tutorial-identifiability-julia} 
    & fix initial conditions, fix some parameters, \par ~~ more outputs, simplify model \\
%    Ebola~\cite[\S 3.6]{chowell2023structural}) & simplify model  
    26 disease models~\cite{dankwa2022structural} & fix some parameters, more outputs
    \\
    Covid~\cite{Tuncer31122022} & fix some parameters
    \\
    Covid~\cite{ciupe2022identifiability} & rescale model
    \\
%    Covid (255 variants)~\cite{massonis2021structural}
    Measles~\cite{turkey-measles} & fix some parameters
    \\
    Seasonal influenza~\cite{nemeth2023structural} & rescale model
    \\
%    \\
    \hline
    \end{tabularx}
    \label{tab:applications}
    \end{table}

Many models in epidemiology are given by ODEs with rational-function right-hand sides (this is the case for all models listed in Table~\ref{tab:applications}).  
%Other techniques in analyzing epidemiology models involve changing the model structure.  
Two recent articles, however, lie outside this setting: one analyzed 
six commonly used growth models that have non-integer power exponents~\cite{growth-models} while the other focused on
age-structured PDE models~\cite{renardy}.  In the first case, the authors reformulated their models using additional state variables so that the resulting models have rational-function right-hand sides and therefore can be readily analyzed by standard methods from differential algebra.  All six models are then found to be structurally identifiable~\cite{growth-models}.
In the second case, the authors extend the differential-algebra framework to handle certain families of PDEs.  
%compare results to the corresponding ODE models.  
They find that the PDE model -- and also the corresponding ODE one -- are structurally identifiable 
when the parameters are constants (that is, not age-dependent) and there is no immigration.  With immigration, however, the PDE model is identifiable while the ODE model is not~\cite{renardy}.  

Another common area of application relates to cancer treatment.  In \cite{eisenberg2017confidence}, models of chemotherapy treatment were found to be structurally identifiable but not practically identifiable. In \cite{t-cell}, the authors examine structural identifiability using steady-state data for a network model involving immune cells, drug, and cancer cells. 
The literature on identifiability using 
steady-state data is limited.
In \cite{saccomani2018union}, the authors analyzed an %oncological 
immunotherapy 
model and found that six of the parameters are unidentifiable, but there are only two degrees of freedom among them; hence, fixing two of them renders the model structurally identifiable.

%that two of the parameters are unidentifiable but fixing them would lead to identifiability.

Other applications include ovarian follicle population dynamics \cite{ballif} and contrast transport models for dynamic constrast-enhanced magnetic resonance imaging (MRI) \cite{transport-model}. The article~\cite{ballif} focuses on a system 
that is
non-autonomous
and so can not be handled by 
the {\tt Julia} package \textit{StucturalIdentifiability}~\cite{julia-software}.  Accordingly, the authors first analyzed
an autonomous submodel, and then used the results to determine that the full model is unidentifiable.  Hence, the authors renormalized certain parameters before investigating practical identifiability. 
In \cite{transport-model}, the authors determined that four (nested) models are structurally identifiable, and subsequently analyzed how practical identifiability of these models depends on the quality of the data. 

%a LTK model and find it is structurally identifiable and then analyze practical identifiability.

These studies motivate several open questions:
%indicate that there are some remaining open questions relating to real-world models in applications:

\begin{question}\label{q:application}
~
\begin{enumerate}
%    \item What families of models have parameters that do not appear in the input-output equations (as in \cite{chowell2023structural})?
    \item What types of parameters are often unidentifiable, and 
    %are common unidentifiable parameters and 
    which parameters should be fixed (assumed to be known) to render the models identifiable (as in \cite{dankwa2022structural, massonis2021structural})?
    \item What families of models are identifiable from steady-state data (as in \cite{t-cell})?
    \item Can the differential algebra approach be extended to handle models that are non-autonomous (as in \cite{ballif}) or involve non-integer power exponents (as in \cite{growth-models})?
\end{enumerate}
    
\end{question}

We pose a final question that leads us to the next section:
What families of unidentifiable models have simple scalings to render the models identifiable (as in \cite{ciupe2022identifiability,nemeth2023structural})?

\section{Identifiable functions and reparametrization}
\label{sec:reparam}
In the prior section, we highlighted several common approaches to handling unidentifiable models in practice.  
% there are a few options for the modeler to remedy this.  
One approach is to
%the modeler can 
adjust the model, for instance, by adding measurements (that is, outputs) to the experimental setup;
this topic is explored further in the next section for the case of linear compartmental models. 
However, if the modeler wants to keep the original model, there is an alternative: one can attempt to reparametrize the model using identifiable functions of the parameters 
so that the resulting reparametrized model is identifiable.  In what follows, we summarize recent progress in this direction.

In some algorithms for computing reparametrizations, we must first find a suitable set of identifiable functions.  
This topic was explored in the context of finding ``identifiable paths/cycles'', which are monomials corresponding to cycles and paths in the graph of a linear compartmental model~\cite{meshkat2014identifiable,BM-2022}. 
More recently, an updated procedure has been developed for finding identifiable functions in nonlinear models \cite{all-identifiable-functinos}.  

Some linear compartmental models can be reparametrized by scaling the state variables~\cite{baaijens2016existence}; necessary and sufficient conditions for such a scaling reparametrization are found in \cite{meshkat2014identifiable}.   
More generally, it was recently shown that linear compartmental models always admit 
 a reparametrization that is a linear function of the state variables, and moreover there are explicit formulas for these reparametrizations in certain ``single-in-single-out" cases 
(one such reparametrization was shown in Example~\ref{example:reparam}) 
 and also 
for models without inputs~\cite{MOS}

Nonlinear models have also seen advances in reparametrizations~\cite{MOS,specialization,FALKENSTEINER2025102385}.  
In \cite{MOS}, a method employing Gr\"obner bases is used to find locally or globally identifiable reparametrizations.  
Another method is presented in \cite{specialization}, which employs specializations of the system to obtain reparametrizations.  The main advantage of the method in \cite{MOS} is that it can always find a globally identifiable reparametrization if one exists, whereas the advantage of the method in \cite{specialization} is that it preserves the structure of the system and is a faster algorithm overall. 
%The step of computing IO equations is the same as in \cite{MOS}, but then the rest is typically faster - the solving to find explicit specialization constraints is much simpler than the solving from \cite{MOS}. The solving step was typically instantaneous in \cite{specialization} unlike in \cite{MOS}, where one sometimes needs to come up with clever ad hoc solutions to speed it up.  
%There is also another method to find globally identifiable reparametrizations \cite{FALKENSTEINER2025102385}.

%Here are some open questions relating to identifiable functions and reparametrizations:
We end this section by posing some related open questions.
\begin{question}\label{q:reparametrize}
~
\begin{enumerate}
    \item What additional families of graphs have easy reparametrizations (e.g., scaling or simple affine form)?
    \item Besides monomials arising from cycles and paths, are there other identifiable functions of parameters that have a graphical interpretation in the model?
\end{enumerate}
\end{question}

\section{Structural identifiability of linear compartmental models} \label{sec:iden-model}
Which linear compartmental models are generically locally identifiable?  
The state-of-the-art (partial) answers to this question are summarized in Tables~\ref{tab:iden}--\ref{tab:uniden}. 
%\ref{tab:unidenparam}.
%~\ref{tab:uniden}, ~\ref{tab:idenparam}, and~\ref{tab:unidenparam}.    
Two specific cases are showcased in the following result, which pertains to 
models 
in which the underlying graph is a bidirected tree~\cite{BGMSS} or a directed cycle~\cite{cycle-cat}:

\begin{theorem}[Bidirected-tree and directed-cycle models~\cite{cycle-cat,BGMSS}] \label{thm:tree-cycle}
~
\begin{enumerate}
    \item A bidirected-tree model with one input and one output is generically locally identifiable if and only if the model has at most $1$ leak, and the distance from the input to output is at most $1$.
    \item A directed-cycle model is generically locally identifiable if and only if it is ``leak-interlacing'' (which essentially means that between any two leaks there is an output or an input, as in Figure~\ref{fig:cycle-4}).
\end{enumerate}
\end{theorem}

\begin{center}

\begin{table}[ht]
\caption{{\bf Identifiable models.}  This table lists classes of linear compartmental models (with at least one input and at least one output) that are generically locally identifiable.  Here, $n$ denotes the number of compartments.
For a {\em directed-path model}, the underlying graph is $1\to 2\to \dots \to n$. For definitions of identifiable-cycle
and identifiable-path/-cycle models, see~\cite{meshkat2014identifiable,BM-2022}.
}
\begin{tabularx}{\textwidth}{X c c c c}
    \hline
Model description & \# inputs & \# outputs & \# leaks & Reference \\
\hline \hline
{\bf Bidirected-tree} with distance \par ~~ from input to output $\leq 1$ 
        & $1$ & $1$ & $\leq 1 $ &  Theorem~\ref{thm:tree-cycle}.1 \\ %\cite[Thm.~5.2]{BGMSS} \\
%Directed-cycle & $1$ & $1$ & $1$ &  \cite[Proposition 5.4]{meshkat2014identifiable} \\
%\hline
{\bf Directed-cycle} with 
    \par
    ~~  interlacing leaks & any & any & any &  Theorem~\ref{thm:tree-cycle}.2\\ %\cite[Thm.~3.6]{cycle-cat} \\
%\hline 
{\bf Directed-path} with $In=\{1\},$ \par ~~ $Out=\{n\}$, $Leak=\{1,n\}$
        & $1$ & $1$ & $2$ &  \cite[Prop.~3.29]{BM-2022} \\
{\bf Identifiable-cycle models} with \par
    ~~ $1\in In$, $1\in Out$,  $Leak \subseteq In \cup Out$
    & any & any & any
    %& $ i \geq 1 $ & $j \geq 1$ & $< i 
    %+j $ 
    &  \cite[Thm.~5]{MSE} %Theorem 5.1 in arXiv version
\\
%\hline
    {\bf Identifiable-path/-cycle models} \par ~~ with
    $Leak=In \cup Out$ & $1$ & 
    %$j \geq 1 $ & $ j+1 $ 
    any & any
    &  \cite[Thm.~3.19]{BM-2022} \\
    %\hline
    {\bf Identifiable submodels} \par ~~ joined by one-way flow
%One-way flow between \\ ~~ identifiable submodels
    & any & any & any &  \cite[Thm.~3.25]{gross2020joining}\\

    \hline
\end{tabularx}
\label{tab:iden}
\end{table}
\end{center}

\begin{center}
    \begin{table}[htb]
\caption{{\bf Unidentifiable models.}  This table lists classes of linear compartmental models (with at least one input and at least one output) that are  unidentifiable.  
A model is {\em strongly connected} (respectively, {\em strongly input-output connected}) if its underlying graph is strongly connected (respectively, is 
connected and every edge is contained in a cycle or in a path from an input to an output).  
See also~\cite[Corollary 3.5]{BGMSS} and~\cite[Theorems~6.2--6.3]{BM-2022}.}
\begin{tabularx}{\textwidth}{X c c c c}
    \hline
Model description & \# inputs & \# outputs & \# leaks & Reference \\
\hline \hline
{\bf Bidirected-tree} 
        & $1$ & $1$ & $\geq 2 $ &  Theorem~\ref{thm:tree-cycle}.1\\ %\cite[Thm.~5.2]{BGMSS} \\
{\bf Bidirected-tree} with distance \par ~~ from input to output $\geq 2$ 
        & $1$ & $1$ & any  &  Theorem~\ref{thm:tree-cycle}.1\\  %\cite[Thm.~5.2]{BGMSS} \\
{\bf Directed-cycle} with \par ~~ non-interlacing leaks & any & any & any &  Theorem~\ref{thm:tree-cycle}.2\\ %\cite[Thm.~3.6]{cycle-cat} \\
{\bf Not identifiable-cycle model} \par
    ~~ when all leaks added & $1$ & $1$ & $1$ &  \cite[Cor.\ 4.2]{BM-2022} \\
{\bf Not identifiable-path/-cycle} \par
~~ {\bf model} when all leaks \par ~~ added, 
strongly connected, 
\par
 ~~ $Leak=In \cup Out$
        & $1$ &
    any & any &
    %$j$ & $j+1$ & 
     \cite[Cor.\ 4.3]{BM-2022} \\

{\bf Strongly connected} & $1$ & any & $> |In \cup Out|$ &  
\cite[Thm.\ 6.1]{BM-2022} \\
{\bf Strongly input-output} %(I-O) 
\par ~~ {\bf connected} 
    & any & $1$ & $> |In \cup Out|$ &  \cite[Thm.\ 6.1]{BM-2022}\\
%Not output-connectable & $i$ & $j$ & $k$ & Cor.\ 3.14 \cite{GHMS} \\
% [ANNE COMMENTED THESE OUT TO AVOID SAYING SOMETHING ABOUT NUMBER OF EDGES, DISTANCE, ETC.]
%Strongly connected without exchange & $1$ & $1$ & $1$ & Thm.\ 6.2 \cite{BM-2022} \\
%Strongly I-O connected w/o edge from in to out & $1$ & $1$ & $2$ & Thm.\ 6.3 \cite{BM-2022} \\
    \hline
\end{tabularx}
\label{tab:uniden}
\end{table}
\end{center}

Table~\ref{tab:iden} tells us some classes of models that have {\em all} parameters (at least) generically locally identifiable.  Nevertheless, in applications, it is desirable to know which parameters are globally identifiable (as opposed to SLING).   
Additionally, it is useful to know which parameters are unidentifiable, as we may need to 
modify the experimental setup (for instance, by adding outputs) or 
reparametrize the model (as described in prior sections).  Recent results in this direction --  classifying parameters as globally identifiable or unidentifable -- are summarized in Tables~\ref{tab:idenparam} and~\ref{tab:unidenparam}, respectively.

%---------
% TABLE 3
%---------
\begin{center}
    \begin{table}[htb]
\caption{{\bf Globally identifiable parameters.}  This table lists classes of linear compartmental models and parameters that are globally identifiable.
For a {\em catenary} model (respectively, {\em mammillary} model), the underlying graph is $1\rightleftarrows 2 \rightleftarrows \dots \rightleftarrows n$ (respectively, is a bidirected star graph with $1$ being the central node).
}
\begin{tabularx}{\textwidth}{X X c}
    \hline
Model description & Parameter(s) & Reference \\
\hline \hline
{\bf Strongly connected} & edge from input to output %(if exists) 
        & \cite[Lem.~2.26]{mammmillary-model} \\
{\bf Catenary} with \par ~~ $In=Out=\{1\}$, up to $1$ leak & all parameters & \cite[\S3.1--3.2]{cobelli-lepschy-romaninjacur} \\
%Catenary models & all parameters & Proposition 6.2 \cite{singularlocus}  \\
{\bf Mammillary} with $In=\{i\}$, \par
 ~~ $Out=\{j\}$, no leaks, and distance \par ~~ from input to output $\leq 1$ 
        & $a_{1j}$ (if $j \neq 1$) and $a_{j1}$ (if $i\neq 1$ or \par ~~ $j \neq 1$) & \cite[Thm.\ 3.2]{mammmillary-model} \\
%Cycle models & $k_{ij}$ depends on structure & Appendix A of \cite{cycle-cat} \\
    \hline
\end{tabularx}
\label{tab:idenparam}
\end{table}
\end{center}

%---------
% TABLE 4
%---------
\begin{center}
    \begin{table}[htb]
\caption{{\bf Unidentifiable parameters.}  This table lists classes of linear compartmental models (with at least one input and at least one output) and parameters that unidentifiable.
For a model with underlying graph~$G$, the {\em output-reachable} subgraph to an output in compartment-$i$ is the induced subgraph of $G$ containing all vertices $j$ for which there is a directed path in $G$ from $j$ to $i$.   
}
\begin{tabularx}{\textwidth}{X X c}
    \hline
Model description & Parameter(s) & Reference \\
\hline \hline
{\bf Any} model, with $j$ a compartment
 \par ~~ that is not in any 
  \par ~~ output-reachable subgraph
%with $|In| \geq 1$ with \par ~~ compartment $j$ such that $j$ is not \par ~~ in any output-reachable subgraph \par ~~ of $G$ and  $j$ is a leak compartment \par~~ or there is a directed edge $j \to l$ \par ~~ out of $j$. 
%Not output-connectable 
        & $a_{0j}$ if $j$ is a leak \par ~~ compartment  or \par ~~ $a_{lj}$ if $j \to l$ is an edge &
        \cite[Cor.\ 3.14]{GHMS} \\
%Identifiable cycle models & all parameters & Theorem 5.1 \cite{MSE} \\
{\bf Identifiable-path/-cycle \par ~~ models} & all parameters &  \cite[Thm.\ 3.19]{BM-2022} \\ %, \par \cite[Thm.\ 5.1]{MSE} \\
    \hline
\end{tabularx}
\label{tab:unidenparam}
\end{table}
\end{center}

Returning to Theorem~\ref{thm:tree-cycle}, the following question arises.

\begin{question}  
\label{q:extend-theorem}
How does Theorem~\ref{thm:tree-cycle}
extend to allow for more than one input and/or output (for bidirected-tree models), or to go beyond bidirected-tree and directed-cycle models?
\end{question}
%In other words, can we characterize identifiability (or unidentifiability) for additional families of graphs?
    %\item Can we find families of graphs that are indistinguishable that are also identifiable?
    %\item What does the singular locus look like for other families of graphs?
\noindent

Question~\ref{q:extend-theorem} can be viewed as asking for identifiability criteria when a graph structure is fixed and the locations of the input(s) and output(s) are varied.  Alternatively, we can ask for identifiability criteria
(necessary or sufficient conditions) when the locations of the input(s) and output(s) are fixed and the graph is varied
%A related problem to {\color{violet} finding necessary conditions for identifiability fixing a graph and varying input/output placement is to find necessary conditions for identifiability fixing the placement of inputs/outputs and varying the graph structure} 
(such criteria could include the
%e.g.,

existence of certain ``exchanges'' or paths, as in~\cite[Theorem 6.2--6.3]{BM-2022}). 

Next, Table~\ref{tab:idenparam} motivates several questions. 
\begin{question}
%First, f
For which models are all parameters globally identifiable?  
More generally, for which classes of models can we readily predict the ``identifiability degree'' -- which is $1$ in the case of globally identifiable models~\cite{cobelli-lepschy-romaninjacur,singularlocus}?
\end{question}
\noindent
On the other hand, which model structures have inherently SLING parameters that cannot be made globally identifiable by adjusting the model, for instance, by adding inputs, outputs, or initial conditions? 

Finally, 
we return to a theme of this article:
%a recent theme pertains to 
adjusting unidentifiable models so that they become identifiable~\cite{BM-2022,MSE}.  

\begin{question}
Given a graph structure, what are the minimal configurations of inputs and outputs that make the model generically locally identifiable (see~\cite{MSE, joubert2018determining})?  
\end{question}

Alternatively, an open problem in this direction is to
predict, based on the model structure, which subsets of the parameters can be removed so that the model becomes identifiable, as in \cite{BM-2022, MSE}.  
Answers would help experimentalists understand the minimal conditions needed so that their experimental setup is suitable for parameter estimation.

\section{Outlook}
\label{sec:end}

Structural identifiability is, in some sense, a solved problem.  A researcher with a (not too large) model can use existing software to check whether the model is structurally identifiable -- and also see which of the parameters are identifiable.  However, in systems biology, we often do not know the precise model.  Which interactions, flows, and degradations  exist, and which need to be included in the model?  How do identifiability properties depend on these answers?  A case study into these questions appears in recent work of Haus, Rengstig, and Thorsen~\cite{haus}, who analyzed more than 3000 (variants of) biomolecular reaction network models for structural identifiability.  A related question concerns how identifiability properties are affected by operations on models~\cite{gogishvili-database,gross2020joining,singularlocus,GHMS}.
Finally, as highlighted in this article, there is a growing literature on theoretical and practical approaches to handling models that are unidentifiable.  This, too, is a promising direction of great practical importance.  
In summary, we see a great synergy between progress made in theory and in applications on the topic of structural identifiability.

 %TC:ignore

\subsection*{Acknowledgements.}
AS was partially supported by the National Science Foundation (DMS-1752672).

\subsubsection*{Declaration of competing interests: None.}
%DECLARATION OF INTEREST: All authors must disclose any financial and personal relationships with other people or organizations that could inappropriately influence (bias) their work. A summary declaration of interest statement is needed in the manuscript file. If there are no interests to declare then please state this: 'Declaration of interest: none'. This summary statement will be ultimately published if the article is accepted. If there are detailed disclosures, they must be included as part of a separate Declaration of Interest form, which forms part of the journal's official records. It is important for potential interests to be declared in both places and that the information matches.

%\subsubsection*{References and recommended reading.}
%In the References, papers of particular interest, published within the period of review, have been highlighted as:

%* of special interest

%** of outstanding interest

\bibliographystyle{plain}

%-------------------
% Commented out, so we can annotate reference
%-------------------
%\bibliographystyle{elsarticle-num} 
\bibliography{bib}

%TC:endignore
\end{document}